\documentclass[twocolumn,superscriptaddress,nofootinbib,aps,prb]{revtex4}

\usepackage{amsmath,amssymb}
\usepackage{MnSymbol}
\usepackage{graphicx}
\usepackage{color}
\usepackage{hyperref}
\usepackage{url}
\usepackage{slashed}
\usepackage{subfigure}
\usepackage{slashed,bbm}
\usepackage{graphics,psfrag,epsfig}
\usepackage{dsfont}
\usepackage{enumerate}
\usepackage{pbox}

\newcommand{\bpsi}{\bar{\psi}}

\renewcommand{\v}[1]{\mathbf{#1}}
\newcommand{\G}{\Gamma}
\newcommand{\la}{\Lambda}

\newcommand{\eps}{\epsilon}
\newcommand{\meas}[1]{\mathrm{d}^2 #1}

\DeclareMathOperator*{\SumInt}{%
\mathchoice%
  {\ooalign{$\displaystyle\sum$\cr\hidewidth$\displaystyle\int$\hidewidth\cr}}
  {\ooalign{\raisebox{.14\height}{\scalebox{.7}{$\textstyle\sum$}}\cr\hidewidth$\textstyle\int$\hidewidth\cr}}
  {\ooalign{\raisebox{.2\height}{\scalebox{.6}{$\scriptstyle\sum$}}\cr$\scriptstyle\int$\cr}}
  {\ooalign{\raisebox{.2\height}{\scalebox{.6}{$\scriptstyle\sum$}}\cr$\scriptstyle\int$\cr}}
}

\begin{document}

\title{Electronic instabilities of the extended Hubbard model on the honeycomb lattice\\
 from functional renormalization}

\author{Yanick Volpez}
\affiliation{Institut f\"ur Theoretische Physik, Universit\"at Heidelberg, Philosophenweg 16, 69120 Heidelberg, Germany}

\author{Daniel D. Scherer}
\affiliation{Niels Bohr Institute, University of Copenhagen, DK-2100 Copenhagen, Denmark}

\author{Michael M. Scherer}
\affiliation{Institut f\"ur Theoretische Physik, Universit\"at Heidelberg, Philosophenweg 16, 69120 Heidelberg, Germany}
\affiliation{Department of Physics, Simon Fraser University, Burnaby, British Columbia, Canada V5A 1S6}

\begin{abstract}
Interacting fermions on the half-filled honeycomb lattice with short-range repulsions have been suggested to host a variety of interesting many-body ground states, e.g., a topological Mott insulator.
A number of recent studies of the spinless case in terms of exact diagonalization, the infinite density matrix renormalization group and the functional renormalization group, however, indicate a suppression of the topological Mott insulating phase in the whole range of interaction parameters.
Here, we complement the previous studies by investigating the quantum many-body instabilities of the physically relevant case of spin-1/2 fermions with onsite, nearest-neighbor and second-nearest-neighbor repulsion.
To this end, we employ the multi-patch functional renormalization group for correlated fermions with refined momentum resolution observing the emergence of an antiferromagnetic spin-density wave and a charge-density wave for dominating onsite and nearest-neighbor repulsions, respectively.
For dominating second-nearest neighbor interaction our results favor an ordering tendency towards a charge-modulated ground state over the topological Mott insulating state. 
The latter evades a stabilization as the leading instability by the additional onsite interaction.
\end{abstract}

\maketitle

\section{Introduction}

In the last years the field of topological states of matter has been subject to stunning progress following the theoretical prediction~\cite{bernevig2006} and experimental observation~\cite{koenig2006} of topological insulators.
A paradigmatic role within this field is played by the quantum anomalous Hall (QAH) state in the Haldane model\cite{haldane1988} -- a model of non-interacting spinless fermions on the half-filled honeycomb lattice giving rise to a quantized Hall conductivity at zero field. 
This model requires an intricate magnetic flux pattern as provided by a complex second-nearest neighbor hopping amplitude $t^\prime$ averaging to zero over the unit cell and breaking time-reversal symmetry.

Later, Raghu \emph{et al.}\cite{raghu2008} and further mean-field studies\cite{weeks2010,dauphin2012,grushin2013} suggested that the QAH state of spinless fermions on the honeycomb lattice can in principle also be obtained dynamically from a large repulsive second-nearest neighbor density-density interaction $V_2$ as
mean-field decoupling of this term generates a ground-state solution with an effective complex second-nearest neighbor hopping $t^\prime$.
Recent elaborate studies in terms of exact diagonalization\cite{jia2013,garcia2013,daghofer2014,laeuchli2015}, infinite density matrix renormalization group\cite{pollmann2015} have addressed the stability of the QAH ground state upon inclusion of quantum fluctuations. 
These studies have found an intricate competition of various phases in the plane of nearest- and second-nearest-neighbor interactions including, e.g., a conventional charge density wave state and a Kekul\'e dimerization pattern\cite{garcia2013,laeuchli2015,pollmann2015}.
In the regime where the topological Mott insulating QAH state was expected, a charge-modulated (CM) density wave phase has been found to be the ground state in Refs.~\onlinecite{jia2013,garcia2013,daghofer2014,laeuchli2015,pollmann2015}.
The QAH state, on the other hand, is completely expelled from the quantum phase diagram.
Note, that numerical evidence for the persistence of the QAH state has been presented in Ref.~\onlinecite{duric2014}.
Further support for the scenario where the QAH is suppressed in the model's phase diagram has been contributed by means of the functional renormalization group within a high-definition resolution of the momentum dependence of the interaction vertex in our Ref.~\onlinecite{scherer2015}.

By now, the spinless model has been investigated in great detail by analytical and numerically exact methods.
At the same time, the physically relevant spin-1/2 case remained elusive to studies operating at the same level of accuracy which can be attributed to the great amount of numerical effort that would be required, e.g. by exact diagonalization techniques.
Further, in the spinless as well as in the spin-1/2 case, an analysis of the problem by means of quantum Monte Carlo (QMC) approaches is typically inhibited by the fermion sign problem.
On the other hand, recent QMC studies\cite{golor2015} have also taken into account non-local density-density interactions on the honeycomb lattice, however, a definite statement about the quantum many-body ground state in the presence of sizable second-nearest neighbor interactions is hindered by limitations on the accessible parameter range.

In the spin-1/2 case, a cousin of the QAH state on the honeycomb -- the quantum spin Hall state -- emerges from an intrinsic spin-orbit coupling as suggested by Kane and Mele\cite{kane2005} in the context of a graphene model. 
The QSH basically corresponds to two copies of the Haldane model with opposite chirality for the two spin projections and is symmetric under time reversal. 
Also here, a next-nearest neighbor repulsion $V_2$ was argued to potentially trigger a phase transition towards the topologically non-trivial QSH state\cite{raghu2008}.
A qualitative difference to the spinless case, however, is the possibility of a local interaction term, i.e., the onsite repulsion $U$, that can in principle affect the balance of the different fluctuations in the particle-hole channel. 

In this work, we address the role of the QSH state in the phase diagram of spin-1/2 fermions on the half-filled honeycomb lattice,
and in particular, whether the onsite repulsion can drive an instability towards the QSH state through the additional energy penalty of charge imbalance throughout the lattice, as it occurs in the CM state.
To that end, we implement the refined multi-patch RG scheme as established in Ref.~\onlinecite{scherer2015} for the spinless case where it supported the absence of a topological QAH state in the phase diagram.
We find that the unbiased inclusion of
fermionic fluctuations of the spin-1/2 fermions exhibits a CM density wave phase with finite wavevector transfer as the leading instability for interaction strength $V_2$ larger than some critical value $V_{2,c}$ depending on the values of onsite and the nearest neighbor interactions.
The QSH state never is found to be the leading instability in the range of interaction parameters where we consider our weak-coupling approach reliable.

The paper is organized as follows: In Sec.~\ref{sec:model}, we introduce the Hamiltonian for the extended Hubbard-model for spin-1/2 fermions on the half-filled honeycomb lattice with short-ranged repulsions. 
In Sec.~\ref{sec:frg}, we present essentials of the fRG method and discuss the multi-patch scheme employed to analyze the leading instabilities of the model. Further, we discuss the convergence of the results with increasing momentum resolution at the example of the critical onsite interaction strength $U_c$ for the quantum phase transition between the semi-metallic state and the antiferromagnetic spin-density-wave state.
Sec.~\ref{sec:res} is then devoted to an analysis of the quantum many-body instabilities of the model within our fRG approach. 
First, we study an exclusive $V_2$ interaction term to compare to the spinless situation. 
Then, we also include onsite and neares-neighbor interactions, $U$ and $V_1$, respectively.
Conclusions are drawn in Sec.~\ref{sec:conclusion} and some technicalities on the functional renormalization group approach can be found in the appendices.

\section{Model}\label{sec:model}

\begin{figure}
\includegraphics[height=0.12\textheight]{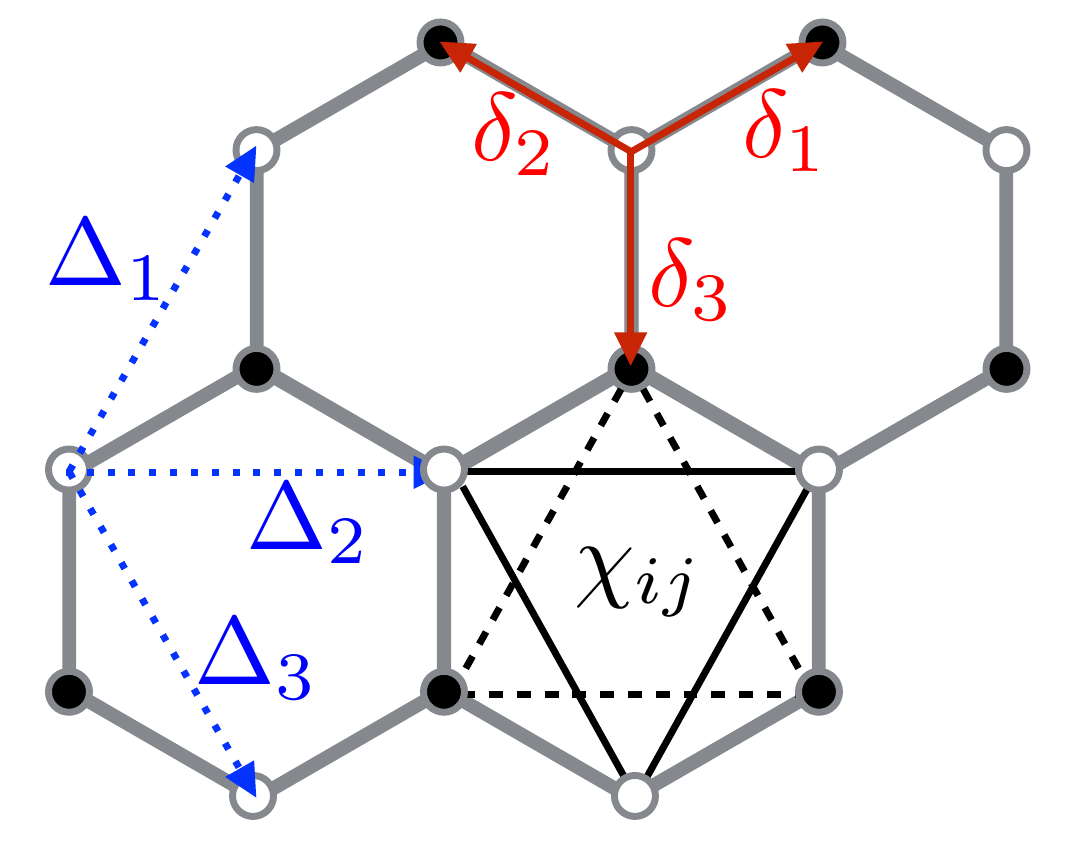}
\hspace{0.8cm}
\includegraphics[height=0.12\textheight]{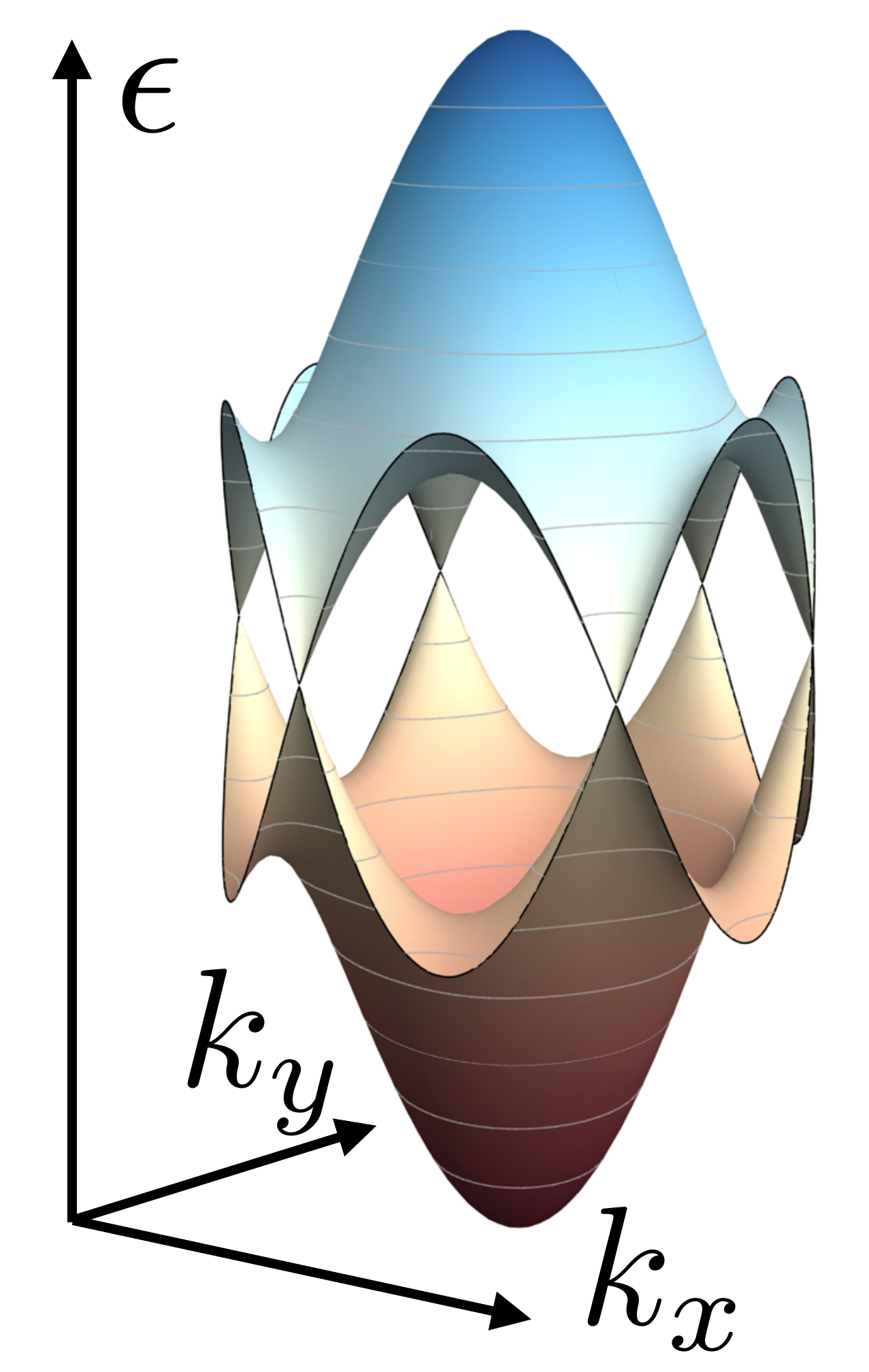}
\caption{Left panel: Lattice structure in real space. The two different sublattices $A$  and $B$ are indicated by empty and filled circles, respectively, connected by the nearest-neighbor vectors $\vec{\delta}_1,\vec{\delta}_2,\vec{\delta}_3$. The second-nearest-neighbor vectors are given by $\vec{\Delta}_1,\vec{\Delta}_2,\vec{\Delta}_3$ and the bond dimerization corresponding to the QAH and QSH ordering patterns is indicated by $\chi_{ij}$, see for example Ref.~\onlinecite{raghu2008} for a discussion of the mean-field order-parameters. Right panel: Nearest-neighbor hopping on the honeycomb lattice provides the depicted energy dispersion with the conduction (blue) and the valence band (red) touching linearly at the $K, K^\prime$ points at the Fermi level.}
\label{fig:lattice}
\end{figure}

We study spin-$1/2$ fermions on the half-filled honeycomb lattice with short-ranged density-density interaction terms.
The Hamiltonian is composed of a single-particle hopping term~$H_0$ and an interaction contribution~$H_{\text{int}}$,
\begin{equation}\label{modelHamiltonian}
 H=H_0+H_{\text{int}}\,,
\end{equation}
where $H_0$ is a tight-binding Hamiltonian
\begin{equation}\label{eq:sp}
 H_0= -t \sum_{\langle i,j \rangle,\sigma} \left( c^{\dagger}_{i,A,\sigma} c_{j,B,\sigma} +\text{h.c.} \right)\,,
\end{equation}
with hopping amplitude $t$ to the nearest-neighbors $\vec{\delta}_1, \vec{\delta}_2, \vec{\delta}_3$ of the hexagonal Bravais lattice with a two-atomic basis labelled by the sublattice index $o \in \{A,B\}$, cf.~Fig.~\ref{fig:lattice}.
The operator $c^{\dagger}_{i,o,\sigma}$ creates an electron at lattice site $i$ in the sublattice $o$ with spin projection $\sigma \in \{\uparrow,\downarrow\}$.

Diagonalization of $H_0$ gives rise to the formation of a spin-degenerate valence and conduction band, touching linearly and isotropically at the $K, K^\prime$ points of the Brillouin zone (BZ). At half-filling, the Fermi surface shrinks to the two Dirac points situated at $K, K^\prime$.
Here, the single-particle density of states vanishes linearly resulting in a semi-metallic behavior of the non-interacting system, Eq.~\eqref{eq:sp}, which is robust against spontaneous symmetry breaking induced by weak interactions.

For the interaction part $H_{\text{int}}$, onsite, nearest-neighbor and second-nearest-neighbor repulsions are taken into account,
\begin{align}\label{Hint}
 H_{\text{int}} &= U \sum_{i,o} n_{i,o,\uparrow} n_{i,o,\downarrow} + V_1 \sum_{\substack{\langle i,j \rangle, \\  \sigma,\sigma'}} n_{i,A,\sigma} n_{j,B,\sigma'} \nonumber \\
 &\quad + V_2 \sum_{\substack{\llangle i,j \rrangle,\\ o, \sigma,\sigma'}} n_{i,o,\sigma} n_{j,o,\sigma'}\,,
\end{align}
where $n_{i,o,\sigma}=c^{\dagger}_{i,o,\sigma} c_{i,o,\sigma}$ represents the local electron density operator
and the sums $\sum_{\langle i,j \rangle}$ and $\sum_{\llangle i,j \rrangle}$ run over nearest neighbors and second-nearest neighbors, respectively. Each pair is counted only once.
Resulting from the inclusion of  these short-ranged repulsions, quantum phase transitions towards different kinds of ordered states occur when the interaction parameters $U, V_1, V_2,...$ exceed critical values,
a subject that has been addressed previously by many different methods\cite{sorella1992,herbut2006,uchoa2007,honerkamp2008,ulybyshev2013,smith2014}.
For example, sizeable pure onsite interactions typically trigger a phase transition towards a fully gapped antiferromagnetic spin-density wave (SDW) state\cite{meng2010,sorella2012,assaad2013} whereas a dominating nearest-neighbor repulsion $V_1$ supports the formation of a charge-density wave (CDW)\cite{khvesh2001,herbut2006,honerkamp2008,wang2014}.
The effect of the second-nearest neighbor interaction $V_2$ is less well-understood even in the absence of other interaction terms.
Mean-field calculations\cite{raghu2008,weeks2010,dauphin2012,grushin2013} suggested the emergence of an interaction-driven topologically non-trivial quantum anomalous Hall (QAH) state in the spinless case and a  time-reversal symmetric quantum spin Hall (QSH) state\cite{raghu2008} in the case of spin 1/2. 
More recent numerical and analytical studies\cite{jia2013,garcia2013,daghofer2014,laeuchli2015,pollmann2015,scherer2015} for the spinless case, however, have found a suppression of the QAH in favor for a charge-modulated (CM) density wave phase with finite wavevector transfer $\vec{K}-\vec{K}^\prime$.
Here, for the case of spin-1/2 fermions, we shall thoroughly investigate the fate of the QSH and CM states upon inclusion of the onsite and nearest-neighbor interactions $U$ and $V_1$.
%

\section{Functional renormalization}\label{sec:frg}

We investigate the quantum many-body instabilities of the model~\eqref{modelHamiltonian} by means of the functional renormalization group (fRG) approach\cite{wetterich1993} for the one-particle irreducible vertices of a correlated fermion systems, see Refs.~\onlinecite{metzner2011,platt2013} for recent reviews.
The multi-patch fRG scheme employed here allows for an unbiased identification of the leading instabilities in the presence of competing correlations\cite{zanchi1998,halboth2000,honerkamp2001,salmhofer2001} by successively integrating out fermion degrees of freedom starting from an initial energy scale $\Lambda_0$ corresponding to the bandwidth down to the infrared $\Lambda\rightarrow 0$.
We now set up the fRG approach in a nutshell and give more details in App.~\ref{app:frg}. 

Consider the fermionic action corresponding to the model Hamiltonian, Eq.~\eqref{modelHamiltonian}, given by
\begin{equation}\label{bareaction}
S[\bpsi,\psi] = -(\bpsi,G_0^{-1} \psi) + V[\bpsi,\psi]\,.
\end{equation}
The first term is the quadratic part with the free propagator
$
G_0(\omega_n,\v{k},b)= 1/(i \omega_n - \epsilon_b(\v{k}))
$
including Matsubara frequency $\omega_n$ and wavevector $\v{k}$. 
Here, we work in the band basis providing diagonalization of the quadratic part of the Hamiltonian $H_0$ with single-particle energies $\epsilon_b(\v{k})$ and band index $b$. The fermionic propagator is diagonal with respect to the spin quantum number.
The interaction term $V[\bar\psi,\psi]$ in Eq.~\eqref{bareaction} is quartic in the fermion fields and can be inferred from the interaction part of the Hamiltonian, Eq.~\eqref{Hint} by introducing Grassmann fields for the operators and transforming  to the band basis in line with the diagonalization of $H_0$. This adds a non-trivial momentum dependence to the coupling function.

In the fRG, the bare propagator is regularized by an infrared momentum cutoff, with energy scale~$\la$,
\begin{equation}\label{eq:momcutoff}
G_0(\omega_n,\v{k},b) \to G_0^{\la}(\omega_n,\v{k},b) = \frac{\theta_{\varepsilon}^{\la}(\epsilon_b(\v{k}))}{i \omega_n - \epsilon_b(\v{k}) }\,.
\end{equation}
Here $\theta_{\varepsilon}^{\la}$ is a smoothened step function with softening length $\varepsilon$ cutting off modes with energies $|\eps_b(\v{k})| \lesssim \la$. 
The modified propagator $G_0^{\la}$ is then used to set up the functional integral representation for the effective action $\Gamma^\la$, which is now scale dependent and generates the one-particle irreducible vertex functions $\Gamma^{(2i)\la}$.
The RG flow is generated upon variation of $\la$ generating a hierarchy of flowing
vertex functions and integration towards the infrared $\la\rightarrow 0$ reproduces the full effective action $\Gamma$, see Apps.~\ref{app:frg} and \ref{app:trunc}.
For our analysis we use a standard truncation that has proven to be suitable for analyzing instabilities in two-dimensional correlated fermion systems.
In this truncation the flow of all $n$-point functions with $n \geq 6$ and also self-energy feedback are neglected.
This amounts to following the RG-scale dependence of the effective interaction vertex $V^{\Lambda}$ which carries a multi-index $k$ gathering Matsubara frequencies $\omega$ as well as wavevectors $\v{k}$ and the band index $b$.
As the most singular part of this quantity comes from the zero Matsubara frequency and we are interested in instabilities, we will also neglect the frequency dependence in the following.
Then, the flow equation of the effective interaction vertex~$V^{\la}$ reads
\begin{equation}\label{eq:flow}
\frac{d }{d \la} V^{\la} = \phi_{\text{pp}} + \phi_{\text{ph,d}}+\phi_{\text{ph,cr}}\,.
\end{equation}
with a contribution $\phi_{\text{pp}}$ from the particle-particle loop, the direct particle-hole loop $\phi_{\text{ph,d}}$ and the crossed particle-hole loop $\phi_{\text{ph,cr}}$. 
All these contributions are bilinears in the scale-dependent vertex function $V^{\la}$ and include a loop-momentum integration. 
We give explicit expressions for the flow equation in App.~\ref{app:flow}.

\subsection{Multi-patch scheme \& momentum resolution}

\begin{figure}[t!]
\includegraphics[width=0.4\columnwidth]{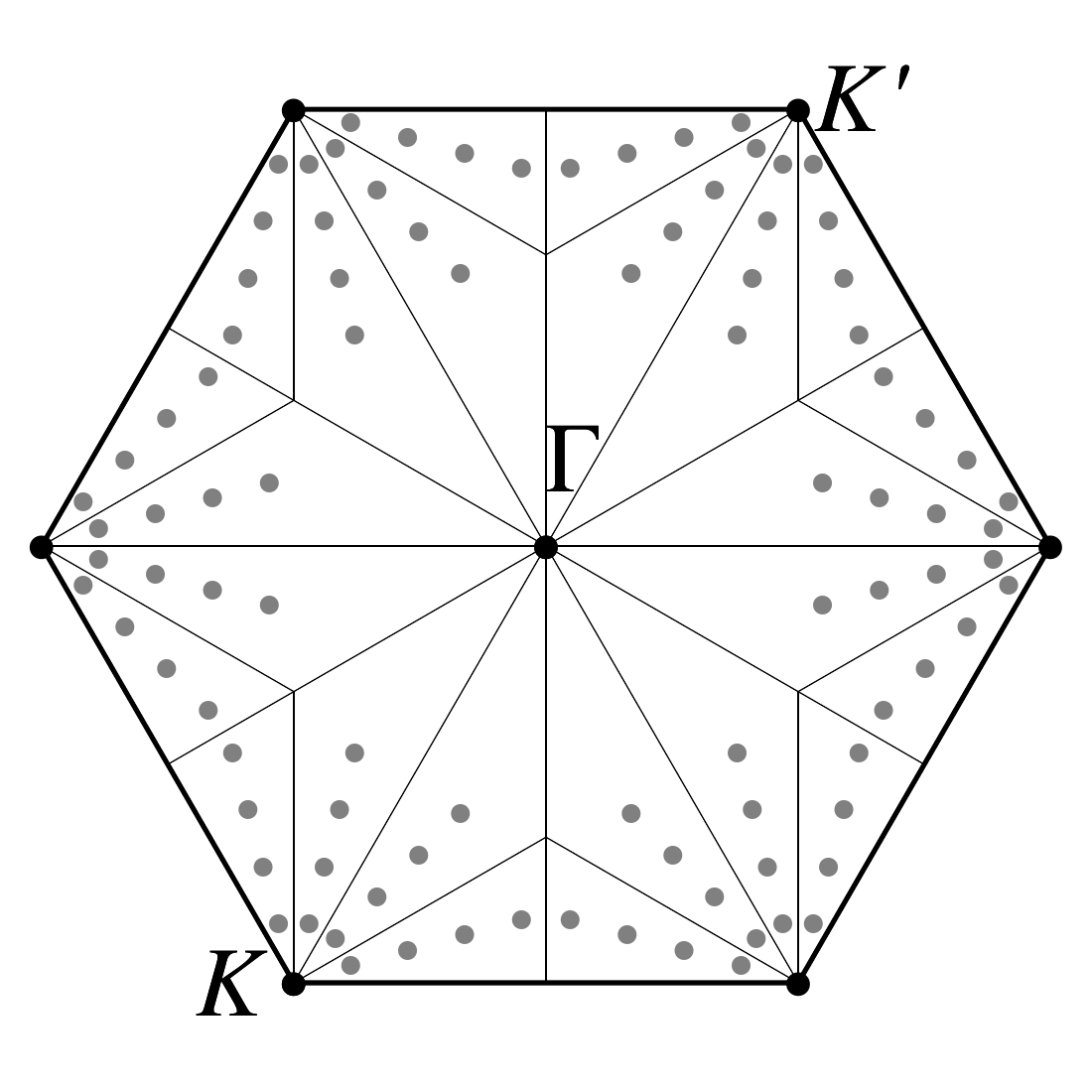}
\includegraphics[width=0.54\columnwidth]{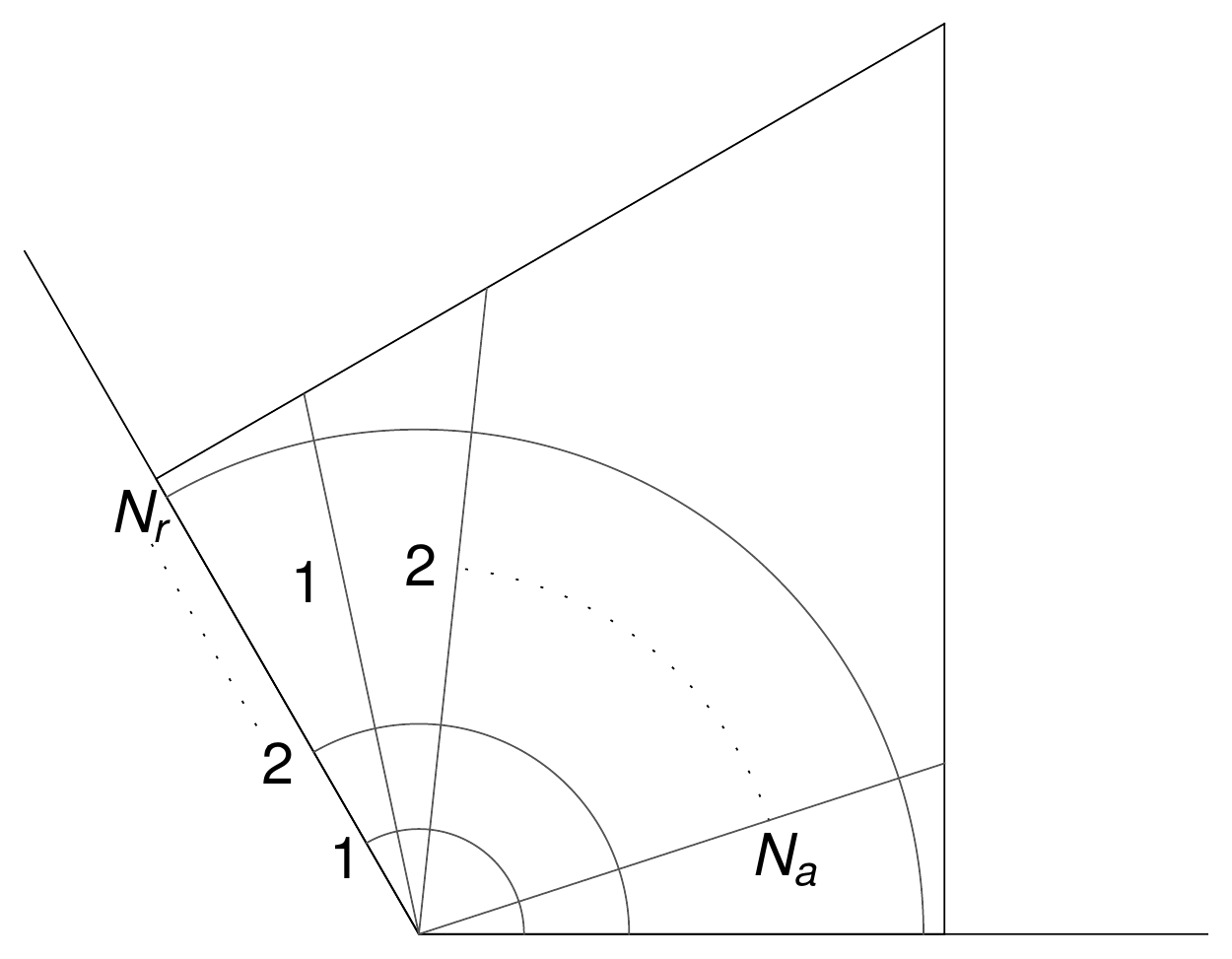}
\caption{Left panel: Sketch of the BZ with patches and patch points with $N_a=4$ and $N_r=4$. Right panel: Numbering of angular and radial patches according to the patching scheme.}
\label{fig:patching}
\end{figure}

The wavevector dependence of the interaction vertex is approximated in a multi-patch scheme.
To this end, the Brillouin zone (BZ) is divided into $N$ patches and each patch is equipped with a representative patch point, see Fig.~\ref{fig:patching} for a pictorial representation.
Then, a given wavevector $\v{k}$ is projected onto its closest patch point $\pi(\v{k})$.
The patching discretization implemented in this work is shown in Fig.~\ref{fig:patching}.
A single patch is composed of all the wavevectors having the smallest distance to the corresponding representative patch point.
We then solve Eq.~\eqref{eq:flow} for the projected vertex function $V^\Lambda(\pi(\v{k}_1),\pi(\v{k}_2),\pi(\v{k}_3),\pi(\v{k}_4))$ which additionally depends on the band indices $b_i$ of the external legs of the four-fermion vertex.
The fourth wavevector is determined by momentum conservation and is subject to an approximation in our scheme by allocating it to its closest patch point.
%
%

For the half-filled honeycomb lattice, the Fermi level is located at the $K,K^\prime$ points where the density of states vanishes linearly.
For the unbiased determination of the instabilities of this model, we have therefore extended the conventional Fermi-surface patching scheme by a refined momentum resolution of the Brillouin zone which additionally resolves momentum dependencies away from the Fermi level, cf.~Fig.~\ref{fig:patching}.
To this end, the patching scheme employed here takes into account several patch rings around the $K,K^\prime$ points, following~Ref.~\onlinecite{scherer2015}.
In the vertex function $V^\la$, each independent momentum argument can then take $N_c\, N_a\, N_r\, N_b$ values, where $N_c=6$ is the number of corners of the Brillouin zone, $N_a$ is the number of angular patches in one corner of the Brillouin zone, $N_r$ of patch rings around the $K,K^\prime$ points and $N_b=2$ is the number of energy bands.
Therefore, the flow Eq.~\eqref{eq:flow} has to be solved for a vertex function with $(N_c\, N_a\, N_r\, N_b)^3 N_b$ components where the fourth band index is not fixed by momentum conservation. 

To evaluate the flow equation, we implement the initial condition $V^{\la_0}$ set at scale $\la_0$ which is given by the two-particle interaction in the bare action, Eq.~\eqref{bareaction}.
The flow equations are integrated out numerically by decreasing the cutoff scale $\Lambda$. 
In flows where the initial interactions $V^{\la_0}$ are large enough, some components of the effective interaction vertex grow large and diverge at a critical
scale $\la_c>0$ indicating a quantum many-body instability.
In practice, the flow has to be stopped at a scale $\la^\ast > \la_c$.
Such a divergence hints at an instability towards a symmetry-broken ground state by a flow to strong coupling and tells us in which channel the leading ordering tendency occurs. 
Further, we obtain a pronounced momentum structure of the vertex function $V^\la$ near the critical scale which is used to extract an effective Hamiltonian for the low-energy degrees of freedom and to determine the leading order parameter.

We note that for the half-filled honeycomb lattice our previous study\cite{scherer2015} has shown that an insufficient wavevector resolution of the Brillouin zone can have an influence on the leading instability which is found as a result of the renormalization group flow, see also Refs.~\onlinecite{wang2012,kiesel2012,kiesel2012b,wang2012b}.
In particular, for the half-filled single-layer honeycomb lattice, this affects the competition between charge-modulated density wave states and the topological Mott insulator.
Here, as in Ref.~\onlinecite{scherer2015}, we overcome this shortcoming by employing the enhanced wavevector resolution of the Brillouin zone, cf. Fig.~\ref{fig:patching}. 
An alternative approach to this problem in terms of the Truncated Unity fRG scheme is presented in Refs.~\onlinecite{liechtenstein2016,pena2016}.

An alternative regularization scheme in terms of the temperature flow\cite{honerkamp2001b} is briefly discussed in App.~\ref{app:temp}. We have cross-checked our results with respect to their sensitivity of the cutoff scheme by evaluating both the momentum and the temperature cutoff and find that our qualitative predictions on the leading instabilities of the extended Hubbard model on the half-filled honeycomb lattice do not depend on the choice of the cutoff.

\section{Analysis of instabilities}\label{sec:res}

Studies of spinless fermions on the half-filled honeycomb lattice with non-local density-density interactions have revealed a competition between various ordering tendencies depending on the relative size of the different interaction parameters $V_1$ and $V_2$. 
Especially, it has been reported that the suggested topological Mott insulator state is destabilized upon inclusion of quantum fluctuations. 
These findings are supported by our recent study within the multi-patch scheme with the refined momentum resolution of the vicinity of the Dirac points, as introduced above. 
In the case of spin-1/2 fermions, the onsite term $U$ adds another source for driving interaction-induced ground states which affects the balance of quantum fluctuations in the phase diagram. While $U$ will primarily drive an antiferromagnetic spin-density wave instability, that is in itself detrimental to the formation of a time-reversal invariant QSH state, it might at the same time suppress charge-ordering tendencies, thus giving way to other momentum structures in the renormalization of the interaction vertex.
For this case, however, there are no calculations for the phase diagram operating at a similar level of reliability as, e.g., the exact diagonalization results due to numerical limitations.
Here, we intend to fill this gap by presenting a high-definition multi-patch RG study of this issue which, despite its approximations, has shown to provide results in agreement with exact methods in large parts of the phase diagram.
Therefore, we map the appearing instabilities for a range of the coupling parameters $U,V_1,V_2$ in units of the hopping amplitude $t$ with $U,V_1,V_2 \in [0,3t]$ which allows to draw tentative phase diagrams of the model.

\subsection{Antiferromagnetic spin-density wave instability}\label{sec:afm}

%
\begin{figure}[t!]
\includegraphics[width=0.95\columnwidth]{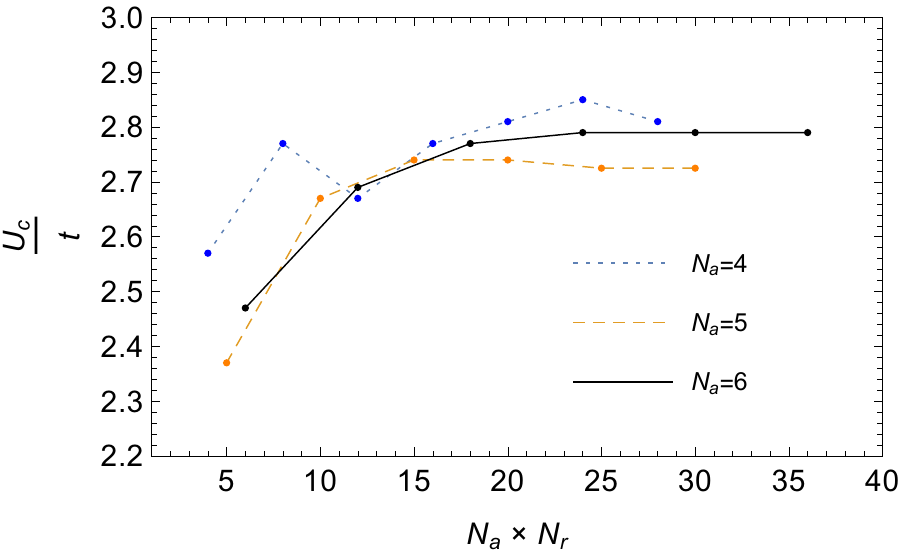}
\caption{Critical interaction strength $U_c$ as a function of the number of patch rings. $N_a$ and $N_r$ denote the number of angular and radial patches, respectively. We observe a smooth convergence in the value for $U_c$ for $N_a=6$. In the following instability analysis we therefore use a patching resolution with $N_\text{tot}=6\,N_a N_r=144$.}
\label{konvergenz_Uc}
\end{figure}
%

The half-filled Hubbard model on the honeycomb lattice, Eq.~\eqref{modelHamiltonian}, with $V_1=V_2=0$ exhibits an instability towards an antiferromagnetic spin-density wave (SDW) state for onsite interactions $U \geq U_c$\cite{}. 
For smaller values of $U<U_c$, the system is in the semi-metallic (SM) phase and does not exhibit any signs of a flow to strong coupling.
We study the dependence of the critical interaction strength $U_c/t$ as a function of the momentum resolution of the patching scheme, cf. Fig.~\ref{konvergenz_Uc}. 
Therefore, we separately vary the number of angular and radial patches, $N_a$ and $N_r$, respectively, and identify the corresponding critical onsite interaction $U_c(N_a,N_r)/t$ for the instability.
For all combinations of $(N_a,N_r)$ and large enough $U/t$, we observe a clear instability towards an antiferromagnetic SDW with a sharply structured leading part of the vertex function in wavevector space which can be cast into the effective Hamiltonian
\begin{align}\label{eq:afsdw}
 H^*_{\text{AF}}=-\frac{1}{\mathcal{N}}\sum_{o,o^\prime}V_{oo^\prime}\epsilon_o\epsilon_{o^\prime} \v{S}_{\v{q}=0}^o\cdot \v{S}_{\v{q}=0}^{o^\prime}\,,
\end{align}
where we have introduced $V_{oo^\prime}>0$, $\epsilon_a=+1$, $\epsilon_b=-1$ and the Fourier transformed spin-1/2 operator is given by $\v{S}_{\v{q}}^o=\frac{1}{2}\sum_{\v{k},\sigma,\sigma^\prime}\boldsymbol{\sigma}_{\sigma \sigma^\prime}c^\dagger_{\v{k}+\v{q},o,\sigma}c_{\v{k},o,\sigma^\prime}$. 
Mean-field decoupling of $H_\text{AF}^\ast$ yields antiferromagnetic spin alignment on the real space lattice.

We observe that for fixed $N_a$ the inclusion of more than one patch ring, $N_r>1$, increases the value for the critical onsite interaction $U_c$ by about 10\%.
Further, for a number of angular patches of $N_a=6$ we obtain a smoothly converging $U_c(6,N_r)$ as $N_r$ is increased. 
The converged value of $U_c\approx U_c(6,4)\approx 2.8\, t$ is reached at $N_r=4$ and a higher $N_r$ does not induce further shifts in $U_c$.
Therefore, the following analysis of instabilities with local and non-local interaction terms is carried out in the patching scheme with $(N_a,N_r)=(6,4)$ which is a trade-off between the accuracy of the momentum resolution and numerical cost. In total this amounts to a momentum resolution of the BZ in $6\,N_a N_r=144$ patches.

Finally, we note that the numerical convergence we have discussed here exclusively refers to the convergence within the specified truncation and approximation scheme.
For the case of a pure onsite interaction $U$ the numerically exact value\cite{meng2010,sorella2012,assaad2013} for the quantum phase transition lies at $U_{c,\text{num}}\approx 3.8\,t$. 
This is considerably larger than our multi-patch fRG value of $U_{c,\text{fRG}}\approx 2.8\,t$ and hints towards an overestimation of fermionic fluctuation effects or the neglect of the fluctuations from collective degrees of freedom within our approach.
Recently, a related high-performance fRG implementation -- the truncated unity fRG (TUfRG)\cite{liechtenstein2016} -- which operates at the same level of approximation to the flow of the vertex function has reported\cite{pena2016} a value of $U_{c,\text{TUfRG}}\approx 3.5\,t$. 
This is considerably higher than our value and can be attributed to heavily increased momentum resolution of the exchange propagators in the TUfRG and suggests a sizeable sensitivity of quantitative results to the momentum resolution.
Further, the TUfRG employs a different regularization scheme -- the $\Omega$ scheme\cite{husemann2009} -- which can also be expected to have an impact on quantitative findings.
For our present purposes, we therefore consider the multi-patch fRG employed here as a tool to qualitatively detect the leading instability of a correlated fermion system in an unbiased way as it treats all the appearing fermionic fluctuations on equal footing
within a infinite-order resummation of all fermionic 1-loop diagrams.

\subsection{Pure next-nearest-neighbor interaction $V_2$}

%
\begin{figure}[t!]
\includegraphics[width=\columnwidth]{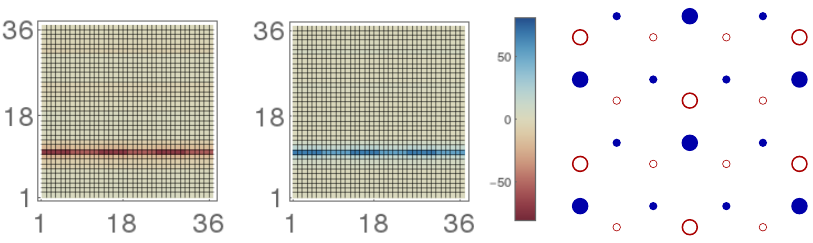}
\caption{Vertex signature for the CM or CDW$_3$ instability. Left and middle panel: Effective coupling in units of $t$ as a function of the incoming patch indices $\pi(\vec{k}_1)$ (vertical) and $\pi(\vec{k}_2)$ (horizontal) with the third patch index $\pi(\vec{k}_3)$ fixed at the patch point with $N_a=1$ in the left lower corner of the BZ, cf.~Fig.~\ref{fig:patching}. For simplicity, the vertex structure is only shown for patch points on the second ring.
The two vertex pictures correspond to different sublattice combinations of in- and outgoing momenta. On the left: $o_1=o_2=o_3=o_4$. In the middle: $o_1=o_3 \neq o_2=o_4$. The calculations were performed with $N_a=6$ and $N_r=4$. The large components in this interaction vertex are used to extract the effective form of the Hamiltonian given in Eq.~\eqref{eq:CDW3}. Right panel: Qualitative charge distribution of the CM mean-field state for $\alpha=\pi/3$.}
\label{V2line}
\end{figure}
%

Next, we study a pure next-nearest-neighbor interaction $V_2\neq 0$ term and setting $U=V_1=0$ as this interaction was suggested to trigger the topological Mott insulator state.
Again, for small values of $V_2$, the semi-metallic behavior from the single-particle picture is stable against interactions.
Beyond a critical value of $V_{2,c} \approx 0.7\,t$, we observe an interaction-driven instability with a sharp momentum structure, cf.~Fig.~\ref{V2line}. 
This indicates a quantum phase transition from the semi-metallic state to an ordered ground state with a charge ordering pattern. 
Explicitly, the sharp momentum structure can be cast into an effective interaction Hamiltonian of the form,
\begin{align}\label{eq:CDW3}
H^*_{\text{CM}} = - \frac{1}{\mathcal{N}} \sum_{o,o'} V_{oo'} \epsilon_o \epsilon_{o'} \left(N_{\v{Q}}^o N_{\v{-Q}}^{o'} + N_{\v{-Q}}^o N_{\v{Q}}^{o'} \right)\,,
\end{align}
again with $V_{oo^\prime}>0$, $\epsilon_a=+1$ and $\epsilon_b=-1$. The density operator reads $N_{\v{q}}^o=\sum_{\v{k},\sigma,\sigma^{\prime}}c^\dagger_{\v{k}+\v{q},o,\sigma}c_{\v{k},o,\sigma^{\prime}}$.
Importantly, the effective Hamiltonian, Eq.~\eqref{eq:CDW3}, exhibits a finite transfer momentum $\v{Q}\approx \v{K}-\v{K}^\prime$.
%
\begin{figure*}[t!]
\includegraphics[width=1.0\textwidth]{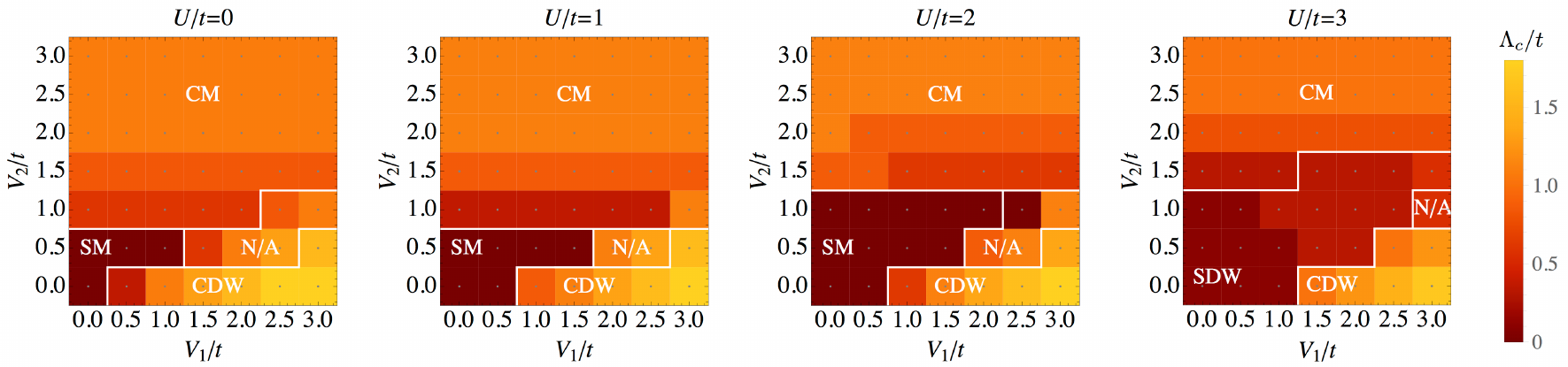}
\caption{Tentative phase diagrams for spin-1/2 fermions on the honeycomb lattice at half-filling as a function of the short-range repulsions $U, V_1$ and $V_2$ in units of the hopping amplitude $t$. The different panels show different values of the onsite interaction $U/t=0$, $U/t=1$, $U/t=2$ and $U/t=3$. The numerical evaluation of the multi-patch RG flow equations was performed with a high-definition patching resolution of $N_a=6$ and $N_r=4$ corresponding to a total number of 144 patches in the BZ. 
In the region marked with N/A, where different ordering patterns or the semi-metallic state meet, a clear identification of the leading instability was not possible as the vertex close to $\Lambda_c$ did not show one clear signature but rather a mix of various instabilities.}\label{fig:uv1v2}
\end{figure*}
%
In real space, the structure of $H_{\text{CM}}^\ast$ corresponds to a infinitely-ranged interaction featuring a modulation with wavevector $\v{Q}$. 
This state is the analogue to the charge-modulated state that has been discussed in detail for the case of spinless fermions\cite{} and has previously also been found in the context of honeycomb bi- and trilayer lattices where it was dubbed the CDW${}_3$ state\cite{mmscherer2012,mmscherer2012b,pena2014}.
The corresponding ground state energy is minimized by the finite complex order parameter $\langle N_{\v{Q}}\rangle =\epsilon_o \Delta_o e^{i\alpha}$ with phase factor $\alpha$ yielding a density modulation $\sim \cos\left(\v{Q}\cdot \v{R}+\alpha\right)$ on the lattice sites. 
A change of $\alpha$ redistributes the charges on the half-filled lattice and at $\alpha=n\pi/3, n\in \mathbb{Z}$ the ground state is minimized by opening a gap, cf.~Fig.~\ref{V2line} for a pictorial representation.
The range of interaction strengths $V_2$ where we find the instability to the CM state extends up to $V_2\approx 4.0\,t$. 
This interaction strength, however, already lies in a regime, where the weak-coupling approximation can be expected to break down as the initial interaction has components that are larger than the bandwidth of the non-interacting system $D=6t$.

We further note that at large $V_2\gtrsim 4.0$ we also find an instability towards the interaction-driven QSH, cf. Ref.~\onlinecite{raghu2008}, with an effective Hamiltonian,
\begin{equation}\label{qsh}
H^*_{\text{QSH}} = -\frac{1}{\mathcal{N}} \sum_{o,o'} V_{o,o'} \epsilon_o \epsilon_{o'} \vec{S}_{f,\v{q}=0}^o \cdot \vec{S}_{f,\v{q}=0}^{o'}
 \end{equation}
where $\vec{S}_{f,\v{q}}^o = \frac{1}{2}\sum_{\sigma,\sigma',\v{k}} f(\v{k}) c^{\dagger}_{\v{k+q},o,\sigma} \vec{\sigma}_{\sigma,\sigma'} c_{\v{k},o,\sigma'}$ with the $f$-wave form factor $f(\v{k}) = \mathrm{sin}(\sqrt{3}ak_x)-2\mathrm{sin}(\frac{\sqrt{3}ak_x}{2})\mathrm{cos}(\frac{3ak_y}{2})$.
At moderate $V_2$, however, this instability only shows up as an artefact of an insufficient momentum resolution of the BZ, e.g., by restricting the multi-patch scheme to a single patch ring close to the Dirac points.
We interpret this finding as an indication for the absence of the QSH state from the phase diagram upon inclusion of sufficiently momentum-resolved fermionic quantum fluctuations.

\subsection{Interplay of short-range repulsions}\label{wehling}

In the next step, we explore the leading instabilities as a function of independently varied interaction parameters $U, V_1$ and $V_2$, cf. Eq.~\eqref{Hint}. To this end, we note that a dominating nearest-neighbor interaction $V_1$ is known to trigger a conventional charge density wave (CDW) beyond a critical $V_{1,c}$ which we find to be  $V_{1,c} \approx 0.3\,t$. We note, that this is close to the critical value obtained from Monte Carlo methods~\cite{wei2014}. This might point to a weak influence of
collective fluctuations close to the phase transition, in accord with the expectation that collective fluctuation effects are less pronounced in the spontaneous breaking of a discrete symmetry. For $V_1 > V_{1,c}$, we observe the CDW instability that can be cast into the effective Hamiltonian
\begin{align}\label{eq:CDW}
H^*_{\text{CDW}} = - \frac{1}{\mathcal{N}} \sum_{o,o'} V_{oo'} \epsilon_o \epsilon_{o'} N_{\v{q}=0}^o N_{\v{q}=0}^{o'}\,,
\end{align}
with the same conventions for  $V_{oo^\prime}, \epsilon_o$ and $N_\v{q}^o$ as above.
This corresponds to a infinitely ranged density-density interaction on the lattice with a sublattice modulation on the $A$ and $B$ sublattices representing a conventional CDW where, e.g., sublattice $A$ is occupied and $B$ is empty or vice versa.

The resulting tentative phase diagrams are shown in Fig.~\ref{fig:uv1v2} as a function of the short-ranged interaction parameters  $V_1$ and $V_2$ for four choices of the onsite interaction $U/t \in \{0,1,2,3\}$. 
Most importantly, we see that for $U\lesssim 2t$, the phase diagram is completely dominated by the semi-metallic state, the CDW state and the CM state. 
Interestingly, for increasing $U$ the critical value $V_{2,c}(U\neq 0)$ is shifted towards larger values indicating a competition between the spin-density wave correlations and the CM correlations, see the panels for $U/t=1$ and $U/t=2$ in Fig.~\ref{fig:uv1v2}.
On the other hand, despite enhanced correlations in the spin channel which support the formation of the QSH state, we do not find an indication for a stabilization of the QSH state as the leading instability due to finite onsite interactions.
In fact, in the region of larger $V_2$, the correlations in the charge channel are always dominant and the CM state turns out to be the leading instability.
For the largest onsite interaction we have explored, i.e. $U=3\,t$, the SDW state reemerges as an instability for small $V_1, V_2$.
Also here, we do not observe a stabilization effect of the topological QSH state due to the onsite interaction for any choice of interaction parameters $U, V_1, V_2$ which is in qualitative agreement with the related study in Ref.~\onlinecite{pena2016}.
We note, however, that our critical scales and interaction parameters turn out different due to the different wavevector resolution and regularization scheme.
In contrast to Ref.~\onlinecite{pena2016}, we do not find a strong suppression of the critical scale at the transition between the SDW state and the CM state to give rise to an intermediate semi-metallic region.
Within our data, however, we can support the statement of a competition between the ordered states due to the increase of the critical value $V_{2,c}(U\neq 0)$ to larger values.
%

\paragraph*{Graphene parameters.}\label{wehling}

In the setup above the short range repulsion parameters $U,V_1,V_2$ are completely independent of each other.
In a real material electrons interact via a (screened) Coulomb potential which leads to a hierarchy in strength of the parameters. 
Using \emph{ab initio} calculations for graphene's interaction parameters within the constrained random phase approximation from Ref.~\onlinecite{wehling2011}, we can provide an estimate for the leading instability for graphene-like systems as a result of short-ranged interaction potentials.
Explicitly, these parameters read $\{U/t,V_1/t,V_2/t\}=\{3.3,2.0,1.5\}$ in units of the hopping amplitude $t\approx 2.8$~eV.
To account for the overestimation of fermionic fluctuations within our approach, cf.~\ref{sec:afm}, we consider variations in the absolute scale of the \emph{ab initio} parameters, maintaining the shape of the potential, i.e. $\{ U/t,V_1/t,V_2/t \} \to \alpha\{U/t,V_1/t,V_2/t\}$ with $\alpha \in [0,1]$.
For $\alpha< \alpha_c\approx 0.75$, we do not observe indications for a flow towards strong coupling as all the vertex components remain finite during the flow for $\Lambda\to 0$. This indicates that the system remains in the semi-metallic state.
In case $\alpha>\alpha_c$ the flow develops an instability towards the antiferromagnetic SDW as given in Eq.~\eqref{eq:afsdw}. 
No other instabilities occur for this choice of parameters.
%
\begin{figure}[t!]
\includegraphics[width=0.9\columnwidth]{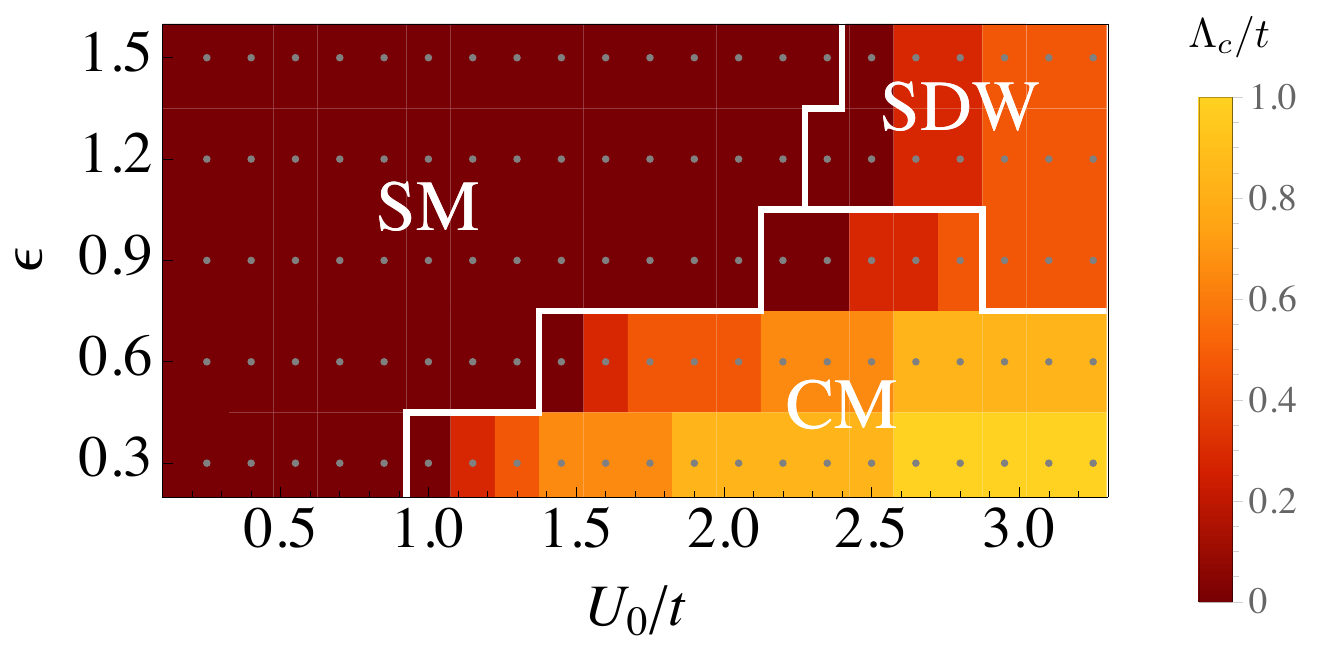}
\caption{Tentative phase diagram for the different sets of short-ranged interaction parameters $U,V_1,V_2$ following a Coulomb-inspired shape of the interaction potential, Eq.~\eqref{eq:pot}, with onsite repulsion strength $U_0$ and screening parameter $\epsilon$. Again, we evaluate the flow equations with a patching resolution of~144 patches in the BZ.}
\label{Uepsilon_phasediagram}
\end{figure}
%

Further, to account for possible modifications of graphene and graphene-like systems affecting the interactions between the electrons, we studied variations of the shape of the screened Coulomb potential in terms of the short-range parameters.
Therefore, we choose an \emph{ad hoc} shape of the potential, reading
\begin{equation}\label{eq:pot}
 U_{\epsilon}(r)= \frac{U_0}{\sqrt{1+\left(\epsilon\, r/a\right)^2}}\,,
\end{equation}
with the lattice constant $a$ and $U_0=U_{\epsilon}(r=0)$ as the strength of the onsite repulsion. 
The parameter $\epsilon$ serves as the screening parameter and in the following we vary $U_0$ and $\epsilon$.
For this choice of the potential, the \emph{ab initio} short-range parameters $U/t,V_1/t$ and $V_2/t$ from Ref.~\onlinecite{wehling2011} are approximately reproduced by $U_0/t\approx 3.3$ and $\epsilon\approx 1.35$.
We note, that we use the suggested form of the potential only for the three short-range interaction parameters and no further interaction terms with $V_n, n>3$ are taken into account in our calculation to account for the finite momentum resolution of the Brillouin zone.
The tentative phase diagram is presented in Fig.~\ref{Uepsilon_phasediagram}. 
Again, the Quantum Spin Hall state is absent, but we find the SDW and CM states next to the semi-metallic phase.
%

\section{Conclusion}\label{sec:conclusion}

We studied interacting spin-1/2 fermions on the honeycomb lattice with repulsive onsite, nearest- and next-nearest neighbor interactions via the functional renormalization group (fRG) for correlated fermions with a focus on the detection of a possible interaction-driven topological Quantum Spin Hall state.
This was motivated through recent studies of interacting spinless fermions by means of exact diagonalization, the infinite density matrix renormalization group and the fRG that supported the suppression of a topological Mott insulating phase in the whole phase diagram of interaction parameters.
Based on the findings in Ref.~\onlinecite{scherer2015}, we used a refined multi-patch RG scheme where fluctuations further from the Fermi surface are considered to overcome the residual influence of an insufficient wavevector dependence on the leading instability in the renormalization group flow.

Here, we complemented these previous studies by investigating the physically relevant case of spin-1/2 fermions additionally featuring onsite interactions which can potentially have an enhancing effect on the correlations in the spin channel supporting the topological QSH state.
For dominating next-nearest neighbor interactions, however, our fRG results favor ordering tendencies towards a charge-modulated ground state over the topological Mott insulating state in agreement with the spinless case.
Further, our results are in qualitative agreement with a recent related fRG study\cite{pena2016} concerning the absence of the topological Mott insulator state in the phase diagram and the appearance of charge-modulated states in the large $V_2$ regime. 
On the other hand, we do not find any sign for an incommensurable charge-modulated state reported in Ref.~\onlinecite{pena2016}. 
This would require a much higher wavevector resolution. This is beyond the applicability of the present multi-patch approach as it requires a much higher numerical cost.
 
To summarize, an independent variation of the interaction parameters for the onsite, nearest-neighbor and next-nearest-neighbor repulsions does not reveal any spot in the tentative weak-coupling phase diagram, where the interaction-driven QSH state represents the leading instability.
Also, for interaction profiles inspired by ab initio parameters for graphene no indication for an topological Mott insulator state is found.
Instead, we identified large parts of the phase diagram where a charge-modulated density wave order is the leading instability and we have found evidence for a competition between the spin correlations and the charge correlations.

\paragraph*{Acknowledgments}

We  thank  C.~Honerkamp, J.~Lichtenstein, D.~S\'anchez de la Pe\~na for useful discussions. D.D.S. acknowledges support by the Villum Foundation. M.M.S. is supported by Grant No.~ERC-AdG-290623 and DFG Grant No.~SCHE~1855/1-1.

\appendix

\section{fRG flow equations}\label{app:frg}

The connected correlation functions of a system of interacting fermions are given by the generating functional for the fully connected correlation functions\cite{negele1988},
\begin{align}
\mathcal{G}[\bar\eta,\eta]=-\ln\int\mathcal{D}\psi\mathcal{D}\bar\psi \,e^{-S[\bpsi,\psi]+(\bar\eta,\psi)+(\bar\psi,\eta)}\,.
\end{align}
In the fRG approach\cite{wetterich1993,metzner2011,platt2013}, we consider the generating functional for the one-particle irreducible (1PI) correlation functions or effective action $\Gamma[\psi,\bar\psi]=(\bar\eta,\psi)+(\bar\psi,\eta)+\mathcal{G}[\bar\eta,\eta]$, which is the Legendre transform $\mathcal{G}[\bar\eta,\eta]$ and the field arguments in $\Gamma$ are given by $\psi=-\partial\mathcal{G}/\partial\bar\eta$ and $\bar\psi=\partial\mathcal{G}/\partial\eta$. 
Note that we use $\psi$ for both, the fields in the micrscopic action as well as for the field arguments of the effective action for notational convenience.

The modification of the microscopic action by means of the regulator function, cf.~Eq.~\eqref{eq:momcutoff}, in the action entering the functional integral yields the scale-dependent effective action~$\G^{\la}$.
The functional flow equation for this version of the effective action is obtained upon the variation of $\G^{\la}$ with respect to $\la$ and reads
\begin{align}\label{RGE}
\frac{\partial}{\partial \la} \G^{\la}[\bpsi,\psi] =& - (\bpsi,\dot{G}_0^{-1} \psi) \nonumber\\
&- \frac{1}{2} \text{Tr} \left( (\mathbf{\dot{G}}_0^{\la})^{-1} \left( \mathbf{\G}^{(2)\la}[\bpsi,\psi] \right) ^{-1} \right)\,,
\end{align}
where ($\mathbf{G_0}^{\la})^{-1}= \mathrm{diag}((G_0^{\la})^{-1},(G_0^{\la t})^{-1})$ and
\begin{equation}
 \mathbf{\G}^{(2)\la}[\bpsi,\psi] = \left( \begin{array}{cc} \frac{\partial^2 \G^{\la}}{\partial \bpsi \partial \psi} & \frac{\partial^2 \G^{\la}}{\partial \bpsi \partial \bpsi}  \\ \frac{\partial^2 \G^{\la}}{\partial \psi \partial \psi} & \frac{\partial^2 \G^{\la}}{\partial \psi \partial \bpsi}   \end{array}  \right)\,.
\end{equation}
The initial condition at the scale $\la_{\text{UV}}$ reads $\G^{\la}_{\text{UV}} = S$, where $\la_{\text{UV}}$ is typically chosen as the bandwidth of the model. 
In the limit $\la \to 0$ one successively integrates out all fermionic fluctuations and obtains the full quantum effective action.

\section{Truncation and approximations}\label{app:trunc}

We expand the effective action $\G^{\la}$ in fields,
\begin{align}
\Gamma^\Lambda[\psi,\bar\psi]&=\sum_{i=0}^{\infty}\frac{(-1)^i}{(i!)^2}\sum_{\substack{k_1,...k_i \\ k_1^\prime,...k_i^\prime}}\Gamma^{(2i)\Lambda}(k_1^\prime,...k_i^\prime,k_1,...k_i)\nonumber\\
&\quad\quad\quad\quad\times\bar\psi(k_1^\prime)...\bar\psi(k_i^\prime)\psi(k_i)...\psi(k_1)\,,
\end{align}
and insert it into the flow equation \eqref{RGE}. Then one obtains an infinite hierarchy of flow equations for the 1PI vertex functions.
To use these equations in applications and integrate the flow equations numerically one has to truncate the tower of equations at a certain level and employ approximations.
For our analysis we follow the RG-scale dependence of the two-particle interaction $\Gamma^{(4)\Lambda}$ only, which carries spin indices $\sigma_i$ and a multi-index $k$ gathering Matsubara frequencies $\omega$ as well as wavevectors $\v{k}$ and the band index $b$.
For our spin-rotation invariant system, we can write the two-particle interaction as
\begin{align}
\Gamma^{(4)\Lambda}_{\sigma_1,\sigma_2,\sigma_3,\sigma_4}=V^\Lambda\delta_{\sigma_1\sigma_3}\delta_{\sigma_2\sigma_4}-V^\Lambda\delta_{\sigma_1\sigma_4}\delta_{\sigma_2\sigma_3}\,,
\end{align}
where we have suppressed the $k_i$ and introduced the effective interaction vertex $V^\Lambda=V^\Lambda(k_1,k_2,k_3,b_4)$.

\section{Flow of the effective interaction vertex}\label{app:flow}

The flow equation for the vertex is given in Eq.~\eqref{eq:flow}
and the particle-particle channel the explicitely reads
\begin{equation}\label{eq:pp}
\phi_{\text{pp}} = \SumInt V^{\la}(k_1,k_2,k,b')L^{\la}(k,q_{\text{pp}})V^{\la}(k,q_{\text{pp}},k_3,b_4)
\end{equation}
with $\SumInt = -A_{\text{BZ}}^{-1}T \sum_{\omega} \int \meas{k} \sum_{b,b'}$.
The direct and crossed particle-hole channels are given by
\begin{align}\label{eq:phd}
\phi_{\text{ph,d}} = &\SumInt [-2V^{\la}(k_1,k,k_3,b')L^{\la}(k,q_d)V^{\la}(q_{\text{d}},k_2,k,b_4) \nonumber \\
 &+ V^{\la}(k,k_1,k_3,b')L^{\la}(k,q_{\text{d}})V^{\la}(q_d,k_2,k,b_4)  \nonumber\\
 &+ V^{\la}(k_1,k,k_3,b')L^{\la}(k,q_{\text{d}})V^{\la}(k_2,q_d,k,b_4)]\,,\\
\phi_{\text{ph,cr}} &= \SumInt V^{\la}(k,k_2,k_3,b')L^{\la}(k,q_{\text{cr}})V^{\la}(k_1,q_{\text{cr}},k,b_4)\,,\label{eq:phc}
\end{align}
and we define $q_{\text{pp}} = -k+k_1+k_2$, $q_{\text{d}}=k+k_1-k_3$ and $q_{\text{cr}}=k+k_2-k_3$. $A_{\text{BZ}}$ denotes the are of the first Brillouin zone and the loop kernel reads
\begin{equation}
L^{\la}(k,k') = \frac{d}{d \la} \left[ G_0^{\la}(k) G_0^{\la}(k') \right]
\end{equation}
with the free propagator $G_0$ due to the neglect of the self-energy.

\section{Flow with temperature cutoff}\label{app:temp}

Instead of the momentum cutoff, cf. Eq.~\eqref{eq:momcutoff}, the temperature $T$ can also serve as a flow parameter\cite{honerkamp2001b}. 
In this scheme, the complete temperature dependence is shifted into the quadratic part of the action by rescaling the fields providing a modified regularized propagator,
\begin{equation}
G_0(\omega_n,\v{k},b) \to G_0^T(\omega_n,\v{k},b) = \frac{T^{1/2}}{i \omega_n - \xi_b(\v{k})}.
\end{equation}
The structure of the flow equations, Eqs.~\eqref{eq:pp},~\eqref{eq:phd} and ~\eqref{eq:phc}, remains the same, for details see Ref.~\onlinecite{honerkamp2001b}.
We used the temperature cutoff scheme to test the reliability of our results upon variation of the cutoff procedure. Our results are stable with respect to variations of the regulator scheme.


\end{document}